# Towards full surface Brillouin zone mapping by coherent multi-photon photoemission

Andi Li, Namitha Ann James, Tianyi Wang, Zehua Wang, and Hrvoje Petek[‡]

[1]Department of Physics and Astronomy and Pittsburgh Quantum Institute, University of Pittsburgh, Pittsburgh, USA

Marcel Reutzel[‡]

I. Physikalisches Institut, Georg-August-Universität Göttingen, Göttingen, Germany

[‡]email: petek@pitt.edu, marcel.reutzel@phys.uni-goettingen.de

*Abstract*

We report a novel approach for coherent multi-photon photoemission band mapping of the entire Brillouin zone with infrared light that is readily implemented in a laboratory setting. We excite a solid state material, Ag(110), with intense femtosecond laser pulses to excite higher-order multi-photon photoemission; angle-resolved electron spectroscopic acquisition records photoemission at large in-plane momenta involving optical transitions from the occupied to unoccupied bands of the sample that otherwise might remain hidden by the photoemission horizon. We propose this as a complementary ultrafast method to time- and angle-resolved two-color, e.g. infrared pump and extreme ultraviolet probe, photoemission spectroscopy, with the advantage of being able to measure and control the coherent electron dynamics.

*Introduction*

Time- and angle-resolved two-photon photoemission spectroscopy (TR-2PP) enables mapping the energy and momentum ($k_{\parallel}$, $k_{\perp}$)-resolved electronic structure and dynamics of the occupied and unoccupied electronic bands of solids [1-3]. Using excitation frequencies from the infrared (IR) to the ultraviolet (UV) range, TR-2PP has been applied to a wide range of condensed matter systems ranging from pristine metals to complex materials and interfaces [4-17]. In TR-2PP spectroscopy, the photon energies $\hbar\omega$ of the pump and the probe laser pulses are chosen such that the pump excites the sample to a real or a virtual intermediate state and the probe induces further upward transition from the excited system to induce photoemission. For photoemission to occur, the combined excitations must impart sufficient energy for the excited electrons to overcome the work function $\phi$ for $k_{\parallel}$ near the surface and bulk Brillouin zone center ($\bar{\Gamma}$–point). To access the entire Brillouin zone, however, photoelectrons must be excited to sufficiently high energies to overcome the photoemission horizon. The photoemission horizon refers to the kinetic energy in the surface parallel motion, $E_{kin}^{\parallel} = m_e \hbar^2 k_{\parallel}^2 / 2$, which cannot do work against the work function because $k_{\parallel}$ of electrons passing through a solid-vacuum interface is conserved [18,19]. To induce two- or multi-photon photoemission (mPP), photoelectrons must absorb energy $m\hbar\omega \geq E_{kin}^{\parallel} + E_B(k_{\parallel}) + \phi$, where $m$ is the photon order, $m_e$ is electron mass, $\hbar$ the reduced Planck's constant, and $E_B(k_{\parallel})$ the in-plane momentum dependent initial state binding energy. Therefore, to map out $E_B(k_{\parallel})$, i.e. the electronic structure and dynamics of unoccupied bands in full surface Brillouin zone, the experiment must overcome the photoemission horizon either by probing with higher $\hbar\omega$ or with higher-order mPP (*cf.* Fig. 1).

While the unperturbed occupied electronic band structure spanning the full Brillouin zone is routinely measured with extreme ultraviolet (XUV) light available at synchrotron facilities or from gas discharge lamps, band mapping of the structure and dynamics of the unoccupied region requires wavelength tunability and ultrafast time resolution. Table-top high-harmonic generation (HHG) sources pumped by femtosecond laser amplifiers can supply ultrafast XUV light pulses with $\hbar\omega>20$eV energy at high repetition rates [20-24]; such laser systems have been applied to perform time- and angle-resolved photoelectron spectroscopy (TR-ARPES), where an intense infrared or optical pump-pulse excites the electronic system, and an XUV-probe pulse interrogates its impact on the electronic band structure of the sample. Remarkable achievements include, for example, probing and switching of exotic properties defined by reduced dimensionality, topological protection, and strong correlation [21,23-32].

In this article, we demonstrate a novel approach to probe deep regions of the Brillouin zone that does not require an XUV light source, and gives access to coherent dynamics at a solid surface. Instead of generating high energy photons in a separate non-linear medium, we take advantage of the non-linear response of the sample to realize a complementary (coherent) approach for band mapping. We overcome the photoemission horizon by exciting higher-order mPP to achieve sufficiently high photoelectron energies to record signal at large in-plane momentum. This enables us to perform surface Brillouin zone mapping of the occupied and unoccupied bands employing only infrared excitation frequencies. More prosaically, we transfer the highly non-linear HHG otherwise used in the XUV light generation directly into the mPP process to potentially access additional information. We show proof of principle experiments by exciting the pristine Ag(110) surface with IR frequencies to detect the occupied Shockley surface band $S_{oc}$, centered at the $\overline{Y}$–point ($k_\parallel \approx 0.7$ Å$^{-1}$) of the surface Brillouin zone, by 4PP via a one-photon resonance with an unoccupied Shockley surface band $S_{un}$.

*Results and discussion*

We elaborate the concept of Brillouin zone mapping by coherent mPP with the example of the pristine Ag(110) surface involving the following scenario [Fig. 1(a)]: An energy slice at $E-E_F=0$ eV shows a schematic 2D anisotropic $k_\parallel=(k_x,k_y)$-resolved band structure of the first surface Brillouin zone of the (110) facet of silver. A Shockley-type occupied surface band $S_{oc}$ exists in the surface projected bulk band gap with a minimum at the $\overline{Y}$–point (brown circle, $k_y \approx 0.7$ Å$^{-1}$). Its electronic and non-linear optical properties have been characterized by static ARPES using 21.2-eV light [33], scanning tunneling spectroscopy [34,35], surface second harmonic spectroscopy (SSHG) [36], inverse photoemission [37,38], and many-body theory [39]. In Fig. 1(a), the maximum accessible in-plane momentum that is expected for $\hbar\omega=21.2$ eV is plotted as a cyan circle; because it encloses the $\overline{Y}$–point, $S_{oc}$ is accessible in a conventional XUV-based ARPES experiment. In mPP with IR light, only the bulk *sp*-band at the $\overline{\Gamma}$–point is accessible in a three-photon process (solid red line, $m=3$), which is the lowest photon order $m$ necessary to overcome the work function $\phi$ of Ag(110); yet, the $\overline{Y}$–point and the $S_{oc}$ band are out of reach. However, the $S_{oc}$ band can be probed by increasing the non-linear order $m$ of the photoemission experiment: In 4PP (dashed red line, $m=4$), the $S_{oc}$ band is in reach.

In the following, we confirm the proposed concept experimentally. We measure energy-, and angle-resolved mPP spectra of the Ag(110) surface at room temperature aligned so that its $\overline{\Gamma Y}$–direction is in the optical plane; details on the ultra-high vacuum system and the optical setup have been reported elsewhere [40,41]. In short, we generate tunable ~20-30 fs laser pulses with a fluence of 1 mJ/cm$^2$ on the sample and a 1MHz repetition rate by a noncollinear optical parametric amplifier (NOPA), which is pumped by a Clark MXR fiber laser oscillator/amplifier system. *p*-polarized light excites the Ag(110) sample at a 45° angle of incidence with respect to the detection axis of the hemispherical electron analyzer. Angle-resolved photoemission spectra are acquired with a 2D delay line detector. The angular acceptance angle of the analyzer is 30°; to report mPP spectra over a broader photoemission angle range, the sample is rotated around the normal axis with respect to the optical plane. We report the final state photoelectron energy, $E_f$, relative to the Fermi level, $E_F$.

Figure 2 shows $E_f(k_y)$-resolved mPP data obtained from the Ag(110) surface; the corresponding excitation diagram for excitation with $\hbar\omega$=1.73-eV photons is shown in Fig. 1(b). In green, we plot the photoemission horizon, as obtained for maximum $E_{kin}^{\parallel}$; electrons excited to energies below this threshold *cannot* be detected unless they undergo momentum scattering to alter their photoemission angle. At the $\bar{\Gamma}$-point, we determine the work function of the Ag(110) surface to be $\phi\approx$4.2 eV, consistent with its reported values [42]. For excitation with 1.73-eV photons, three photons (*m*=3) are sufficient to excite photoelectrons above the vacuum level at the $\bar{\Gamma}$-point; the Fermi edge is mapped to $E_f \approx$5.2 eV. In the next higher order of photoemission, i.e. in 4PP, a replica of the Fermi edge is detected at $E_f \approx$6.9 eV. $\hbar\omega$-dependent measurements confirm the classification of those photoemission spectral features to three- and four-photon processes, respectively [Fig. 3]. By contrast, at the $\bar{Y}$-point ($k_y\approx$0.7 Å$^{-1}$), three-photon absorption lifts the $S_{oc}$ band to above $\phi$, but still below the photoemission horizon, thus preventing its detection. Nevertheless, in 4PP, the $S_{oc}$ band is raised above the photoemission horizon to $E_f \approx$6.8 eV [Fig. 2(a)], making it thus detectable in the mPP experiment. The $\hbar\omega$-dependent spectroscopy clearly identifies $S_{oc}$ band feature being photoemitted in a four-photon process [Fig. 3]. We extract the binding energy of the $S_{oc}$ band at the $\bar{Y}$-point to $E_B \approx$0.1 eV with respect to $E_F$, which agrees with XUV-based ARPES experiments ($E_B^{lit} = 0.1$ eV) [33]; minor deviations from the literature value can be attributed to the near-resonant excitation with spectrally broad femtosecond laser pulses. The excitation wavelength is selected to take advantage of a one-photon (near-)resonance condition with the unoccupied Shockley surface band, $S_{un}$ [*cf.* excitation diagram in Fig. 1(b)]. The 4PP process is thus enhanced by the $S_{un} \leftarrow S_{oc}$ transition even though the $S_{un}$ band does not appear as a distinct feature in the 4PP experiment in Fig. 2(a). The resonance between the Shockley surface bands is consistent with its enhancement of SSHG on the Ag(110) surface [36]. Thus, the occupied and unoccupied bands that are otherwise hidden below the photoemission horizon, become accessible in higher-order mPP when exploiting the non-linear response of the sample.

The mPP data presented in Fig. 2(a) thus demonstrates that the non-linear energy conversion process typically applied in separate non-linear media to generate light of higher frequency can similarly be directly excited in a sample and detected through the mPP process. To illustrate the resemblance of both concepts, in the following, we first frequency double the infrared laser pulses in a BBO crystal outside the photoemission apparatus to obtain 3.50-eV photons. Accordingly, $E_{kin}^{\parallel}$ and the maximum reachable in-plane momentum range widen; the $\overline{Y}$-point becomes accessible in a two-photon process [*cf.* blue circle Fig. 1(a), *m*=2]. In 2PP [Fig. 2(b)], the $S_{oc}$ band minimum is detected at $E_f \approx 6.9$ eV, corresponding to a binding energy of $E_B \approx 0.1$ eV with respect to $E_F$, which is in agreement with the 4PP measurement. Accordingly, a slope of two is obtained for the plot of $E_f$ for $S_{oc}$ in $\hbar\omega$-dependent measurements [Fig. 3]. We note, that the additional strongly dispersive bands in Fig. 2(b), which are less prominent though detected in 4PP in Fig. 2(a), are not known and will be discussed in a separate manuscript.

We further point out that the concept of the photoemission horizon and thus the necessity of sufficiently large kinetic energy photoelectrons to probe large $k_\parallel$ is a pure solid-state effect arising from space-periodic arrangement of the lattice atoms. This makes, for example, above-threshold multi-photon photoemission (ATP) [19,40,41,43,44], a phenomenon accompanying mPP driven with IR frequencies, a two-dimensional problem. Conventionally, for mPP excited at the $\bar{\Gamma}$ –point, photoelectrons absorbing more photons than necessary to overcome the work function are treated as ATP. With increasing $k_\parallel$, however, an electron populating a final state may not have sufficient energy in the surface direction to overcome the photoemission horizon. In our mPP data reported for Ag(110), this becomes obvious in Fig. 2(a), where 3PP at the $\bar{\Gamma}$–point is sufficient to be detected and thus the replica structure in 4PP can be attributed to ATP; by contrast, at the $\bar{Y}$–point, 4PP is not an above threshold process but rather the lowest order of photoemission. These considerations stand in contrast to ATP from nanostructured materials, where band momentum integrated processes obscure the origin of the photoelectrons and thus their in-plane momentum information. Instead, the local field enhancements determine the photoexcitation physics [45]. Similarly, as well the highly non-linear excitation of atoms and molecules, where the concept of above threshold ionization was initially developed [46], is not affected by the considerations discussed here, because $k_\parallel$ is not a conserved quantity.

Finally, we emphasize that the proposed methodology is applicable to probe coherent ultrafast phenomena in the full surface Brillouin zone. For example, interferometrically time-resolved multi-photon photoemission techniques [2,5,6,40,47-50] are applicable to those electronic bands. Moreover, it opens a new approach for performing measurements where the excitation field can both drive dipole transitions and interband ballistic electron acceleration in solids [51-53]. Similarly, field induced light band structure engineering is especially promising at those long wavelengths and high electric field strength applied in our work [54,55].

*Conclusion*

In conclusion, we report a coherent approach to perform photoemission spectroscopy, and measure the ultrafast electron dynamics of electronic bands extending deep into the Brillouin zone. Instead of providing large energy photons in the process of HHG, as typically done in optical-pump-XUV-probe TR-ARPES experiments, intense infrared light can be used to induce multi-photon processes leading to photoemission spectral features above the photoemission horizon. Proof of principle data obtained on the Ag(110) surface presented in this article thus provokes the application of mPP to novel materials whose physical and chemical properties are defined by their electronic band structure at the Brillouin zone edges, such as Dirac materials, or transition metal dichalcogenides. Thereby, higher-order mPP yield can be enhanced by resonant driving of selected dipole transitions [41], as well as by excitation at characteristic frequencies of the materials dielectric function, e.g. at the epsilon-near-zero condition, where the materials response is shown to be dominantly non-linear [17,56]. We thus anticipate that time- and angle-resolved photoemission data from such materials will become increasingly accessible in typical laser-based photoemission laboratories without having to resort to more complex XUV pulse generation. Furthermore, the presented methodology can be straightforwardly extended to novel photoelectron detection schemes, opening, for example, the door towards interferometrically time-resolved multi-photon momentum microscopy [57,58].

*Acknowledgements*

The authors gratefully acknowledge financial support from DOE-BES Division of Chemical Sciences, Geosciences, and Biosciences Grant No. DE-SC0002313. M.R. acknowledges continuous support from Stefan Mathias, as well as funding by the Alexander von Humboldt Foundation.

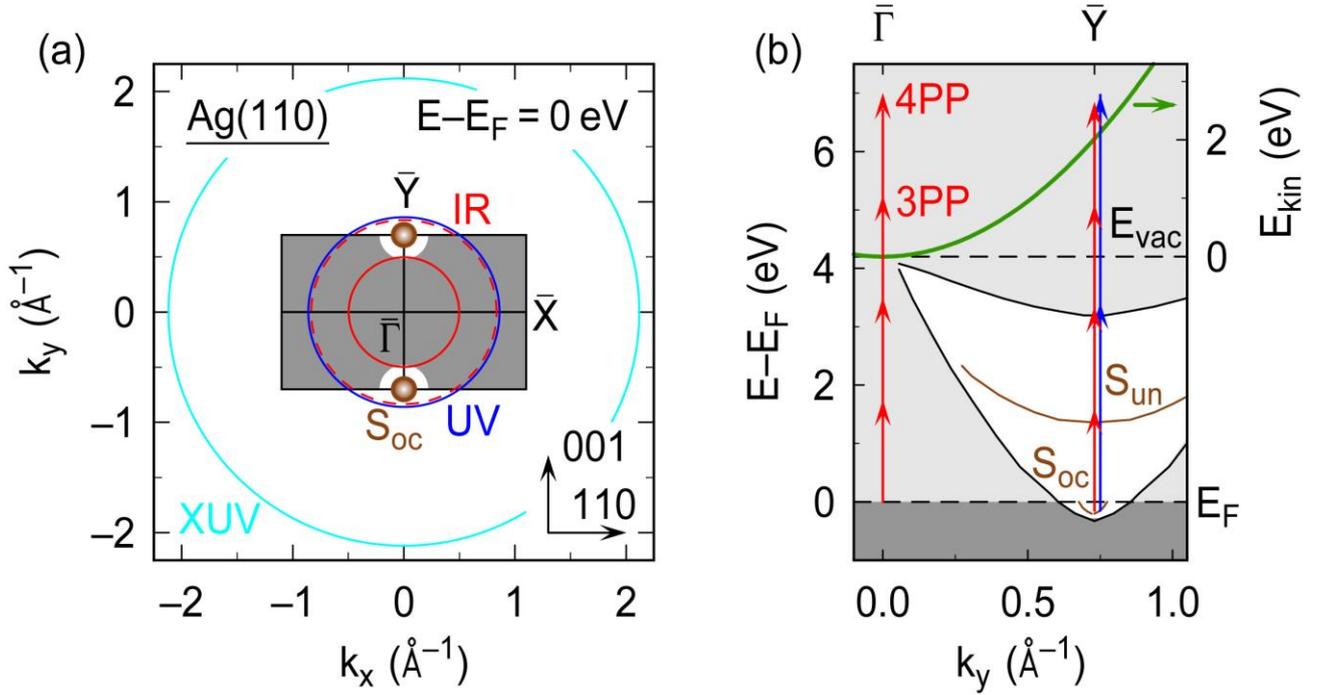

**Figure 1** Accessible in-plane momentum range within higher-order mPP spectroscopy. (a) Two-dimensional cut through the surface projected Brillouin zone of Ag(110) at the Fermi energy; the grey region represents surface projected occupied bulk bands of the first Brillouin zone. At the $\bar{Y}$-point, an occupied Shockley type surface band ($S_{oc}$, brown circle) is found in the surface projected band gap (white filled semicircle). The red circles depict the maximum accessible in-plane momentum when excited with IR photons ($\hbar\omega$=1.73 eV) in 3PP (solid line) and 4PP (dashed line); the blue and the cyan circles illustrate 2PP and 1PP with UV ($\hbar\omega$=3.50 eV) and XUV ($\hbar\omega$=21.2 eV, *cf.* Ref. [33]) light, respectively. (b) $E_f(k_y)$-resolved excitation diagram of Ag(110) cut along the $\overline{\Gamma Y}$-direction. Red and blue arrows indicate excitation pathways excited with 1.73-eV and 3.50-eV photons, respectively. In the surface projected band gap (white), an occupied, $S_{oc}$, and an unoccupied, $S_{un}$, surface band are found. The green parabola indicates the photoemission horizon. Dark and light grey areas depict occupied and unoccupied surface projected bulk bands, respectively.

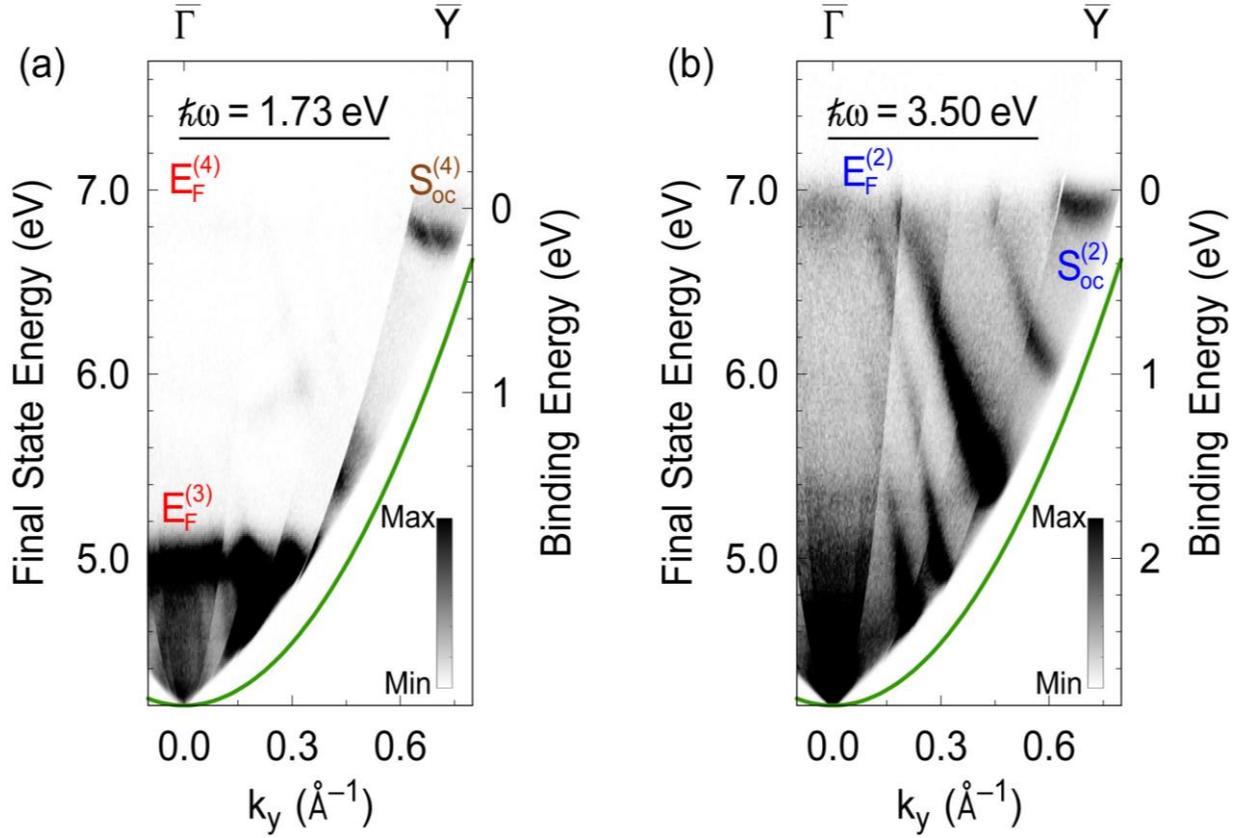

**Figure 2** mPP spectroscopy of Ag(110). (a) $E_f(k_y)$-resolved mPP spectra excited with 1.73-eV photons. At the $\bar{\Gamma}$-point, the Fermi edge, $E_F^{(m)}$, is probed for $m = 3$ and 4. At the $\bar{Y}$-point, the occupied $S_{oc}$ band is only accessible in 4PP, labelled $S_{oc}^{(4)}$. Note that the binding energy axis is only appropriate for the 4PP processes. (b) A comparable binding energy region can be investigated in 2PP with 3.50-eV photon excitation. The Fermi edge and the $S_{oc}$ band are probed in two-photon processes at the $\bar{\Gamma}$- and the $\bar{Y}$-points, respectively. The photoemission horizon is indicated by the green parabola. Discontinuous jumps in the mPP intensity are caused by the assembly of spectra from multiple $E_f(k_y)$-measurements for different sample rotation angles with respect to the electron analyzer.

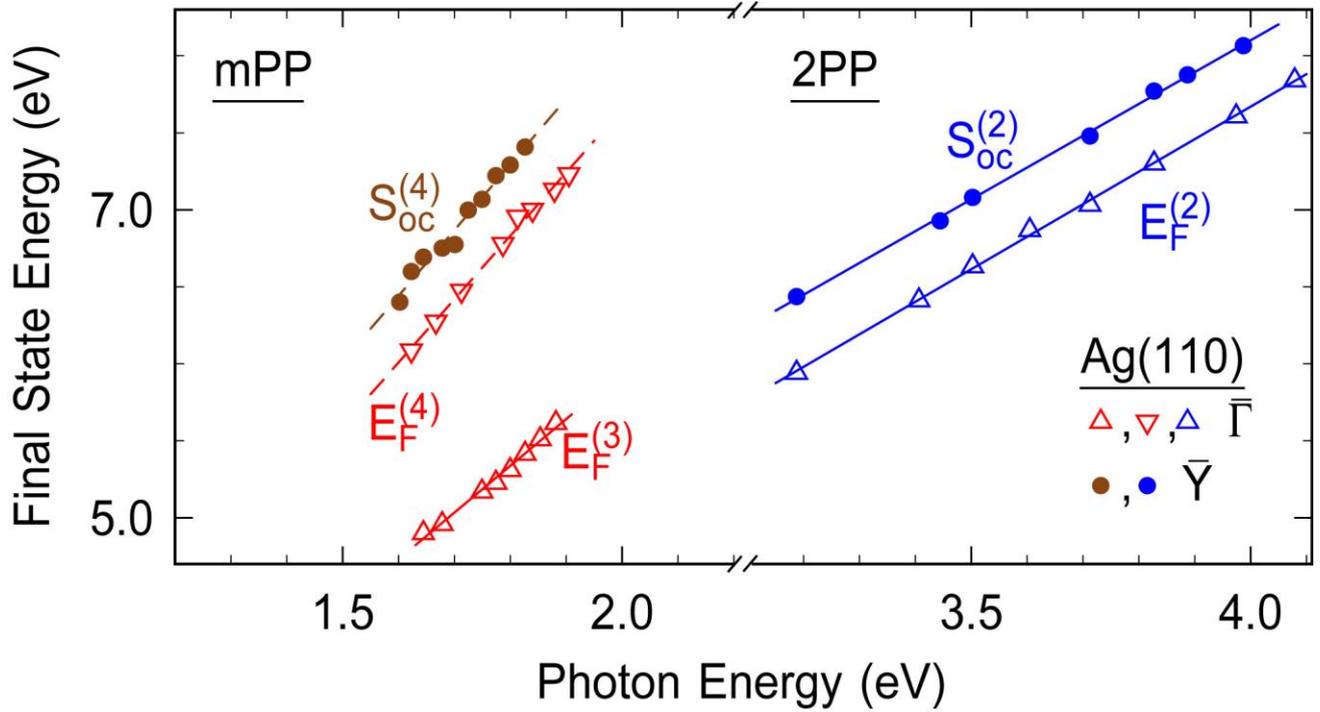

**Figure 3** Photon energy dependent $E_f$-energies of the occupied Shockley band, $S_{oc}^{(m)}$, and the Fermi edge, $E_F^{(m)}$, recorded at the $\bar{Y}$-point, and the $\bar{\Gamma}$-point, respectively. The Fermi edge shifts for $m$=3, 4, 2 with slopes of 3.0, 4.1, 2.1, respectively. The $S_{oc}$ band slopes for $m$=4 and 2 are 4.3 and 2.1, respectively, as expected for coherent mPP processes. The $E_F^{(4)}$- and $E_F^{(2)}$-data are offset by -0.4 eV on the $E_f$-axis for better illustration.